# An optical corrector design that allows a fixed telescope to access a large region of the sky

E.F. Borra, G. Moretto, and M. Wang

Centre d'Optique, Photonique et Laser - COPL
Département de Physique - Université Laval, Québec, PQ, Canada G1K7P4

**Abstract.** We discuss a family of two-mirror correctors that can greatly extend the field accessible to a fixed telescope such as a liquid mirror telescope. The performance of the corrector is remarkable since it gives excellent images in patches contained within a field greater than 45 degrees. We argue that this performance makes fixed telescopes competitive with tiltable ones.

## 1. Introduction

In the Early 1980s (Hewitt 1980) the astronomical community considered building optical telescopes having diameters as large as 30 meters; but the original enthusiasm was soon tempered by the mechanical, optical and financial challenges that would be faced. Some unusual designs were considered but abandoned (e.g., Barr 1980). The design of a telescope is considerably simpler, and its cost lower, if it only points to the zenith. However, its field of view with a conventional corrector is limited to a few degrees, diminishing its usefulness. The usefulness of fixed telescopes obviously increases with their accessible field and it is worthwhile to seek designs that allow to observe far from the optical axis of the primary. If a practical corrector allowing to access a sufficiently large field can be found, it will make fixed telescopes competitive.

The corrector is particularly important for liquid mirror telescopes since they cannot be tilted. The original suggestion by Borra (1982) that modern technology renders liquid mirror telescopes practical has led to the demonstration of a diffraction limited 1.5-m mirror by Borra et al (1992), an article that also gives a wealth of technological details, followed by a 2.5-m mirror (Borra, Content & Girard 1993). Following observations with rudimentary 1-m and a 1.2-m LMTs (Content et al. 1989), an astronomical observatory housing a 2.7-m f/2 liquid mirror has been built that saw first light in late 1992 and has now been operated for an observing season (Hickson et al 1994), giving images with FWHM < 2 arcseconds, compatible with the seeing expected at a sea level site. Liquid mirrors are interesting in other fields of science besides Astronomy. For example, a lidar facility at the University of Western Ontario housing a 2.6-m diameter liquid mirror receiver has been built and is in routine operation: access to a larger region of the sky is also desirable for lidar work.

A landmark paper by Richardson & Morbey (1987) has shown that if one only corrects over a CCD-sized field, it is possible to compensate with warped auxiliary mirrors the aberrations of a parabolic mirror observing 7.5 degrees off-axis. While their design is not practical, it has the merit of showing that correction can be done. Recently, Borra (1993) has explored analytically the fundamental limits within which one can correct the aberrations of a parabolic mirror observing at a large angle from the zenith, showing that the aberrations can, in principle, be corrected to zenith distances as high as 45 degrees. However, this was only a theoretical exploration and practical corrector designs must still be demonstrated.

In a recent paper (Wang, Moretto, Borra & Lemaître 1994), we have considered a simple one-mirror corrector design that uses the active mirror technology pioneered by Lemaître (1989). We now have built and tested a small prototype mirror that demonstrates that the basic idea works (Moretto et al. 1994). This practical design gives adequate performance for spectroscopy but has insufficient image quality and field of view for imagery.

In this article, we discuss a family of 2-mirror correctors that give images and fields of view usable for imagery. The limited aim of this article is merely to demonstrate the existence of a practical design and we did not try to optimize it.

## 2. The BMW Corrector

We have studied a two-mirror corrector design that we have whimsically dubbed the BMW corrector, an acronym made from the initials of the names of the authors that also happens to be the same as the one of a known make of high-performance automobiles. In this article we only present the general features of the corrector, along with a few examples. We find that the design is versatile and that, for a given zenith distance, there is a rich variety of possible configurations, yielding very different focal lengths and geometrical setups. For example, it is possible to place the tertiary mirror close to the secondary, to yield a compact system or, alternatively, place it far from the secondary but near the ground, for easier access. It will take some effort to explore, document and classify all the possible solutions. This will be left to a future article.

The history of two-mirror correctors has been briefly reviewed by Baker (1969) and Robb (1978). Paul (1935) has shown that two spherical mirrors can be located to correct the spherical aberration and coma of a parabolic primary. The design was improved by Baker (1969) who showed that it is possible to correct all third order aberrations as well as fifth order spherical aberration with a two-mirror system having a slightly turned down edge. To the basic Paul-Baker design, we have added an aspheric shape by applying two-dimensional 10th order polynomial corrections to the spherical surfaces of secondary and tertiary and introduced additional degrees of freedom by allowing decentering and tilting of the secondary and tertiary mirrors.

We thus have anamorphic aspheric secondary and tertiary mirrors. These are aspheric surfaces with bilateral symmetry given by

$$Z = \frac{C_x x^2 + C_y y^2}{1 + \{1 - (1+K_x)C_x^2 x^2 - (1+K_y)C_y^2 y^2\}^{1/2}}$$
$$+ A_r\{(1-A_p)x^2 + (1+A_p)y^2\}^2$$
$$+ B_r\{(1-B_p)x^2 + (1+B_p)y^2\}^3$$
$$+ C_r\{(1-C_p)x^2 + (1+C_p)y^2\}^5$$

where Z is the sag of the surface parallel to the Z axis, $C_x = 1/R_x$ & $C_y = 1/R_y$, are the curvatures in X and Y directions, $K_x$ & $K_y$, are the conic coefficients in X and Y, $A_r$, $B_r$, $C_r$ & $D_r$, are the rotationally symmetric portions of 4th, 6th, 8th, and 10th order polynomial deviations from the conic and $A_p$, $B_p$, $C_p$ & $D_p$, represent the non-rotationally symmetric components of the 4th, 6th, 8th, and 10th order deformation from the conic.

For large zenith distances, the additional degrees of freedom and the asphericity help to control higher order aberrations. The resulting surfaces of revolution are larger than the primary mirror, giving unacceptable vignetting. However, by using only the off-axis segments that actually collect the light rays, we obtain secondary and tertiary mirrors having reasonable diameters.

By imposing that the third order aberrations be zero as well as a plane focal surface, we obtained analytical solutions extending the discussion of the Paul-Baker corrector in Schroeder (1987) to which we added decenter and tilt of the secondary and tertiary. The analytical relations, valid for any conic constant but not including the polynomial correction, give initial solutions for the design parameters that are fed to the optimization routines of the ray-tracing software. We use the well-known commercial optical design software CODE V. The discussion in this article is restricted to polynomial correction so that our solutions are probably not the best ones possible.

### 3. Correctors for a 4-m Primary

We have selected as an example a 4-m diameter f/6 parabolic primary. We imposed correctors yielding effective focal lengths of 24 meters, which give a good sampling of the PSF with a typical CCD working in the best seeing conditions on earth. It is possible to obtain designs having similar performance for other much larger, or smaller, focal lengths. We wish to stress that the designs presented here are probably not the best possible ones. The purpose of this paper is simply to introduce the concept and show that practical designs yielding good images are possible. Detailed exploration of the design is in progress and will require considerable additional effort.

Figure (1) shows a schematic of a 4-m mirror and BMW correctors observing at 7.5, 15 and 22.5 degrees from the zenith. The optical axes of the correctors and the optical axis of the primary are contained in the same plane. The designs place the tertiaries near the optical axis, yielding a practical configuration. In particular, the inset in Fig. (1) shows that all tertiaries are near each other. With such a design only the secondary would have to move substantially to access a different region of sky, on a curved rail of a polar-coordinates mount, while the tertiary mirror would move very little. If we picture the tertiary and detector as constituting a small altazimuth telescope that collects the light from the secondary mirror, Fig. (1) shows something approximating a small telescope in three altitude positions. The geometry of this "telescope" is not quite constant but with additional design work we can probably obtain configurations equivalent to the same small telescope oriented at different altitude angles, albeit with different shapes of the mirrors and tilts of the focal planes. Nevertheless, even without this additional optimization, we have arrived to a very practical geometry.

Figure (2) shows the spot diagrams for point sources observed at 14.9, 15.0 and 15.1 degrees from the zenith and for point sources displaced from the same positions by 0.1 degrees in the orthogonal direction. The parameters of the corrector have been optimized to give good images simultaneously for the 6 spots; in other words, a single corrector gives the 6 images. It is possible to obtain better images by optimizing the design over a smaller field. For a 12 arcminute field and a 4-m primary, the diameters of respectively the secondaries and tertiaries are 1.6 and 0.6 meters for the corrector at 7.5 and 15 degrees, and 1.5 and 0.5 meters for the corrector at 22.5 degrees. The tertiaries are small but the secondaries are a bit large, although manageable. We feel, based on our present experience with this corrector, that with additional design work it should be possible to obtain solutions with smaller secondaries. This expectation is buttressed by the fact that, due to an increased effort at finding smaller mirrors, the secondary and tertiary of the corrector working at 22.5 degrees are actually smaller than those of the other two. Also, polynomial corrections may not yield the ideal shapes: We merely selected them because they were a handy option available from the software. For the same reason, there certainly are solutions yielding more compact correctors.

Figure (1) shows some vignetting from the tertiary mirror and the structure needed to support it but it is tolerable since the tertiary is significantly smaller than the primary. The focal surface is at the edges of, but within, the beam from the secondary, causing some additional vignetting. It can be minimized to tolerable values by using a 45-degree mirror to deflect the light to a side detector. With additional work, it may be possible to locate the focal surface totally out of the beam.

Table (1) gives the 100% and 80% encircled energy diameters for the PSFs from the 3 correctors, the root mean square (RMS) and peak to valley (P-V) deviations of the wavefronts obtained from a 36 Zernicke polynomial fit and the RMS spot diameter. Tables (2) and (3) summarize the parameters of the

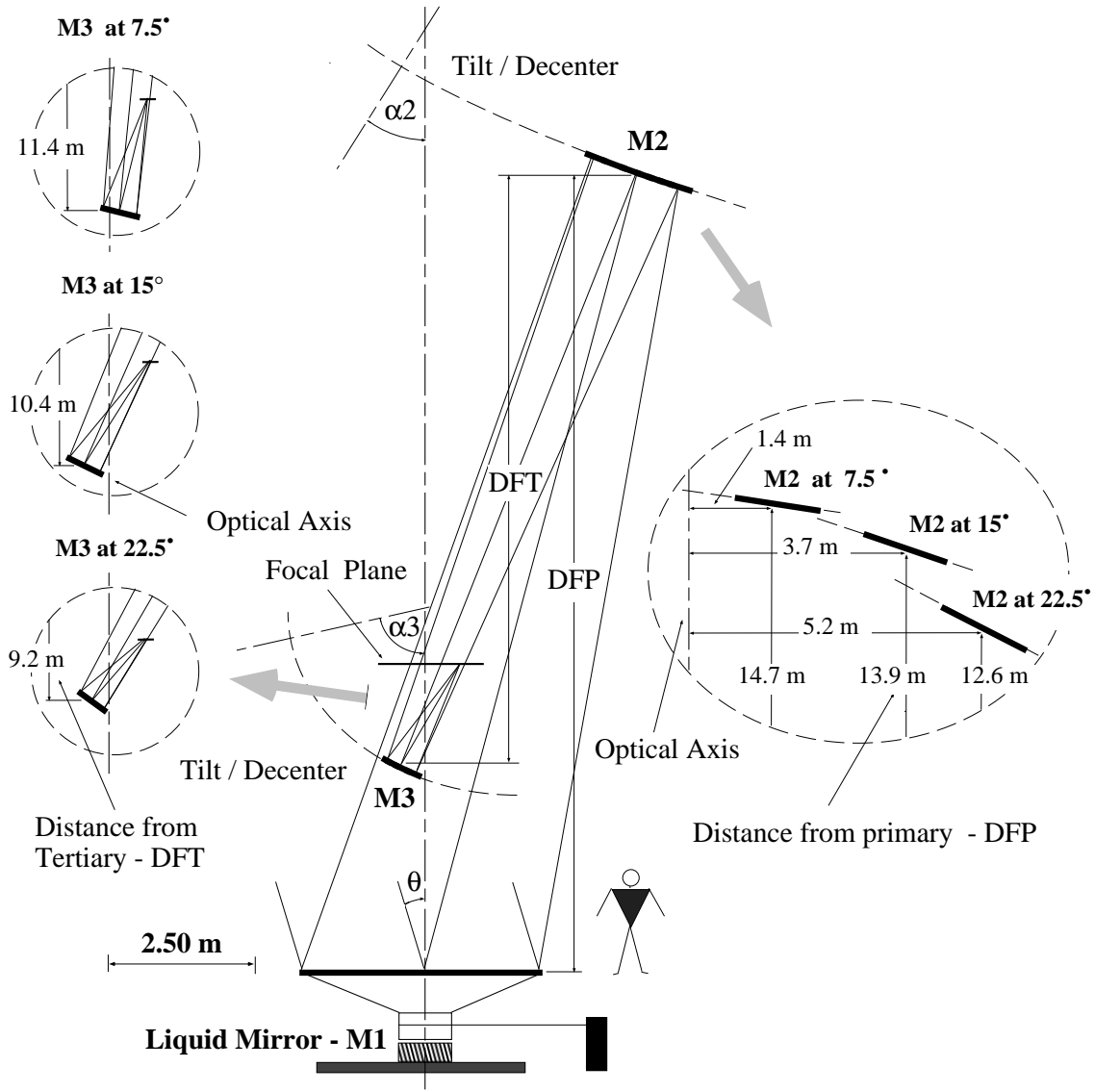

**Fig. 1.** It shows a schematic design of a 4-m diameter f/6 mirror and 3 BMW correctors observing at 7.5, 15 and 22.5 degrees from the zenith.

anamorphic secondaries and tertiariers and the parameters of the correctors for zenith distances of 7.5, 15 and 22.5 degrees

To illustrate the versatility of the design, we have calculated a more compact corrector for the same 4-m primary. We show it in Figure (3), where the telescope observes at 15 degrees from the zenith. The spot diagrams are very similar to those shown in Figure (2) and have very similar encircled energy values.

## 4. Discussion

The BMW corrector gives good images at large zenith distances but can the telescope track? The simplest system consists of a driftscanning survey telescope using a rigid corrector tracking electronically with a CCD detector. The information is stored and the nightly observations coadded with a computer. Imagery with a fixed driftscanning telescope has been demonstrated by McGraw, Cawson and Keane (1986) and by Hickson et al. (1994) with a liquid mirror telescope, and slitless spec-troscopy by Schmidt, Gunn & Schneider (1987). The optical and mechanical setups are simple since the corrector is set for a particular zenith distance and does not have to be adjusted to work at different zenith distances. As suggested by Content (1992) a corrector designed for a given zenith distance $\theta$ (e.g., 22.5 degrees) can be used to observe objects passing anywhere within a field of view of $2\theta$ (e.g., 45 degrees) by moving it at different azimuths but at a fixed zenith distance, as shown in Figure (4). This is the only degree of freedom allowed, everything else (e.g., mirror shapes and distances) are fixed.

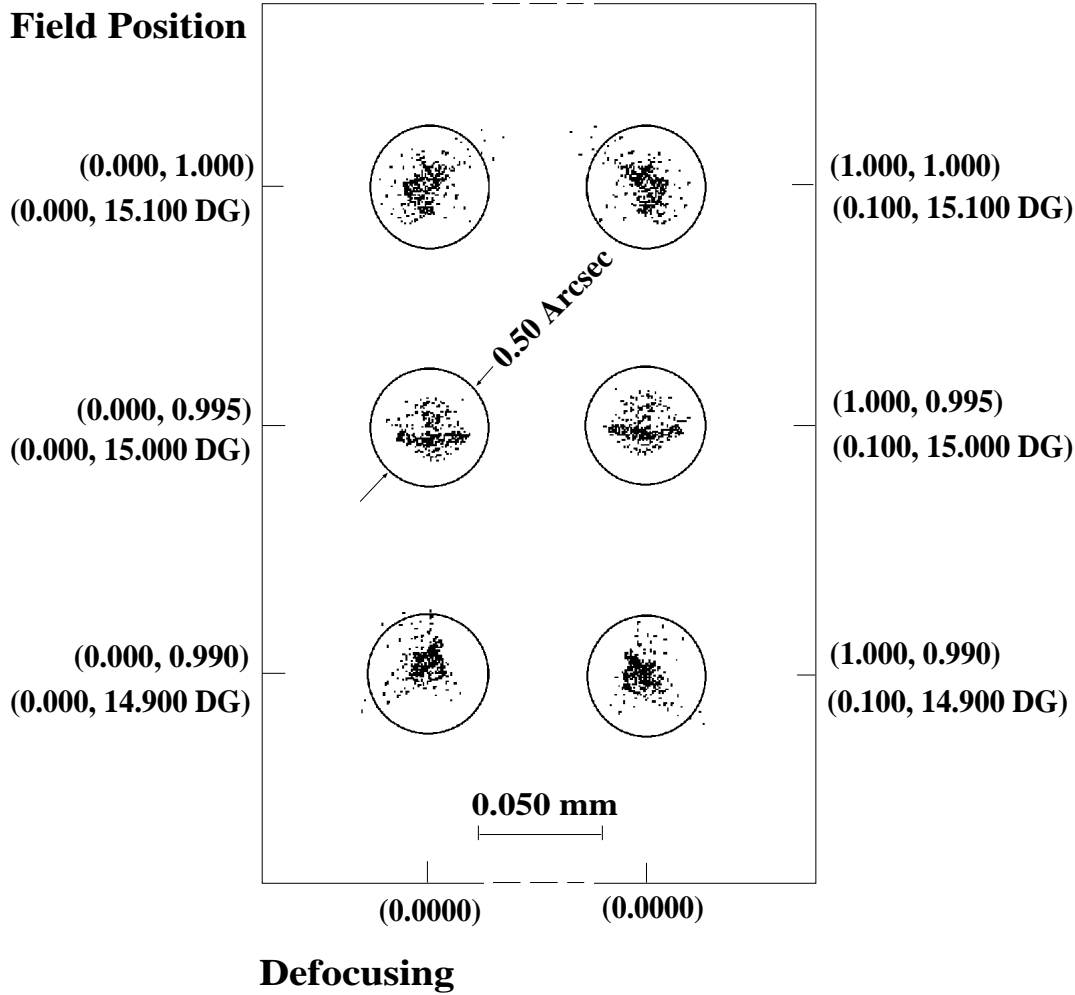

**Fig. 2.** It shows the spot diagrams for point sources observed at 14.9, 15.0 and 15.1 degrees from the zenith as well as 3 spots displaced by 0.1 degrees from those in the orthogonal direction. The 6 spots are given by the same corrector. The inches have diameters equal to 0.5 arcseconds.

A more versatile, and complex, system could track by moving and warping the mirrors to follow an object in the sky. The geometry of the corrector and the shapes of the mirrors must change in real time to track. With a configuration similar to the one in Fig. (1) only the secondary would have to move substantially during tracking, on a polar-coordinates mount, while the tertiary mirror and detector would move very little. The tertiary and detector can be thought as constituting a small altazimuth telescope that collects the light from the secondary mirror.

Prima facie, a f/6 telescope seems slow -although the overall length of the setup is 14-meters-, requiring a tall structure; which is however not as critical for a fixed telescope as it is for a tiltable one since the shelter merely consists of a tall silo. A local architect estimated that a 20-m tall silo having a diameter of 8-m, would cost about $US 120,000, indicating that the cost penalty of a slow telescope is not serious. We have not estimated the cost of the frame of the telescope but this is also less critical since it always remains vertical.

Table (3) shows that the shapes of the secondary and tertiary mirrors are not simple, since these anamorphic aspherics are highly aspherical off-axis segments having different radii of curvature in the y and x directions, and one must find practical ways to make them. A possible labor-saving technology would use a warping harness, polish a mirror to a sphere that then relaxes to the required shape (e.g., Nelson et al. 1980). Alternatively, one could grind and polish with a computer-controlled tool. If the telescope is required to track in real time, one shall have to warp the mirrors in their cells.

**Table 1.** Parameters that characterize the correction at selected zenith angles: a single two-mirror corrector corrects at $\theta \pm 0.1$

| $Field(x,y)$ [Degrees] | $RMS^a$ [Waves] | $P$-$V^b$ [Waves] | 100 % $EED^c$ [Arcseconds] | 80 % $EED^c$ [Arcseconds] | $RMS$-$SD^d$ [Arcseconds] |
|---|---|---|---|---|---|
| (0.0, 7.4) | 0.284 | 1.584 | 0.284 | 0.167 | 0.131 |
| (0.0, 7.5) | 0.188 | 0.997 | 0.198 | 0.121 | 0.096 |
| (0.0, 7.6) | 0.267 | 1.460 | 0.255 | 0.163 | 0.130 |
| (0.1, 7.4) | 0.280 | 1.556 | 0.282 | 0.163 | 0.128 |
| (0.1, 7.5) | 0.189 | 0.977 | 0.196 | 0.122 | 0.097 |
| (0.1, 7.6) | 0.263 | 1.448 | 0.248 | 0.163 | 0.128 |
| (0.0, 14.9) | 0.369 | 2.047 | 0.570 | 0.182 | 0.172 |
| (0.0, 15.0) | 0.342 | 2.051 | 0.357 | 0.233 | 0.176 |
| (0.0, 15.1) | 0.391 | 2.547 | 0.580 | 0.228 | 0.204 |
| (0.1, 14.9) | 0.363 | 1.931 | 0.580 | 0.179 | 0.166 |
| (0.1, 15.0) | 0.351 | 2.064 | 0.357 | 0.241 | 0.181 |
| (0.1, 15.1) | 0.376 | 2.519 | 0.570 | 0.214 | 0.197 |
| (0.0, 22.4) | 0.541 | 3.899 | 0.826 | 0.405 | 0.332 |
| (0.0, 22.5) | 0.279 | 1.663 | 0.569 | 0.222 | 0.181 |
| (0.0, 22.6) | 0.486 | 2.692 | 0.836 | 0.433 | 0.329 |
| (0.1, 22.4) | 0.483 | 3.566 | 0.799 | 0.388 | 0.312 |
| (0.1, 22.5) | 0.308 | 1.721 | 0.612 | 0.257 | 0.208 |
| (0.1, 22.6) | 0.379 | 2.312 | 0.909 | 0.358 | 0.290 |

[a] Root Mean Square deviations of the wavefront at a wavelength of 6320 Å.
[b] Peak to Valley deviations of the wavefront at a wavelength of 6320 Å.
[c] Encircled Energy Diameter.
[d] RMS spot diameter

**Table 2.** Geometrical parameters.

| $\theta$ [Degrees] | Surface | Radius [mm] | Distance [mm] | Displacement and Tilt (DAR) | | | |
|---|---|---|---|---|---|---|---|
| | | | | $x$ [mm] | $y$ [mm] | $z$ [mm] | $\alpha$ [Degrees] |
| 7.50 $\pm 0.1$ | STOP | -48000.0 | -16316.74 | 0.00 | 0.00 | 0.00 | 0.00 |
| | 2 | Table 3 | 11288.11 | -23.17 | -4108.31 | 0.00 | -22.95 |
| | 3 | Table 3 | -258.50 | -1.25 | -2757.26 | -216.23 | -67.52 |
| | IMAGE | INFINITY | 0.00 | 0.00 | 0.00 | 0.00 | 0.00 |
| 15.0 $\pm 0.1$ | STOP | -48000.0 | -15605.29 | 0.00 | 0.00 | 0.00 | 0.00 |
| | 2 | Table 3 | 10288.75 | 1.19 | -143.20 | 0.00 | -33.48 |
| | 3 | Table 3 | -0.10 | 7.89 | -2360.81 | -130.00 | -74.59 |
| | IMAGE | INFINITY | 0.00 | 0.00 | 0.00 | 0.00 | 0.00 |
| 22.50 $\pm 0.1$ | STOP | -48000.0 | -14007.69 | 0.00 | 0.00 | 0.00 | 0.00 |
| | 2 | Table 3 | 11045.58 | -3.86 | 255.26 | 0.00 | -68.87 |
| | 3 | Table 3 | -0.10 | 6.00 | -2463.07 | -909.23 | -78.15 |
| | IMAGE | INFINITY | 0.00 | 0.00 | 0.00 | 0.00 | 0.00 |

To assess the applicability of warping, one must consider the maximum amplitude of the deflection, the stress in the material and the applicability of the theory. The maximum stress in a mirror is particularly important since it must not exceed the elastic limit of the material; otherwise the mirror does not spring back. To estimate the stresses in a mirror we model it with a circular plate supported around the edges and subjected to a uniform load since this yields an analytical solution for the maximum deflection $w_{max}$ at the center of the plate. The load q is then given by (Timoshenko & Woinowsky-Krieger 1959)

$$q = \frac{16Eh^3 w_{max}}{3(1-v)(5+v)a^4} \qquad (1)$$

where $v$ is Poisson's ratio, h the thickness of the plate, E Young's modulus and a the radius of the plate.

The stress in the plate is then given by

$$\sigma_{max} = \frac{3(3+v)qa^2}{8h^2} \qquad (2)$$

We have determined the maximum amplitude of deflexions for the mirrors of the three correctors with respect to their best fitting spheres. In the case of the secondary computed at 7.5 degrees, we obtain a maximum amplitude of 67.8 *mm* after considering the difference of the curvatures in the x and y directions and adding the polynomial corrections. If we consider the secondary of the 7.5 degrees corrector for a thickness of 25 mm, corresponding to a reasonable aspect ratio of 64, we find $\sigma_{max} = 96.6\ Kg\ mm^{-2}$, a large value for glass but less than the maximum stress of AISI 420 steel (120 $Kg\ mm^{-2}$) that is used to make metallic mirrors. The deflection of the tertiary exceeds the maximum stress. For the correctors at angles greater than 7.5 degrees, the stresses of the deflections with respect to the best-fitting spheres are greater than the maximum stress of AISI 420 steel and one could not use the elastic deformation method to generate them *from a sphere*. The mirrors could of course be generated by a combination of elastic deformations and computer generated cutting and polishing. For example, one could generate an intermediate shape with a computer driven tool and bend it in its cell during tracking. As a matter of fact, an elliptic surface would greatly alleviate the problem; unfortunately we cannot compute this since the elastic theory only exists for spherical surfaces. As discussed earlier, if the telescope driftscans, only one shape is needed that can be generated with a computer driven tool. We also expect to find, with more computer work, solutions with lower deflections, albeit with different geometries and plate scales. This will be explored in our continuing investigations of correctors for fixed telescopes.

**Table 3.** Geometrical parameters of the anamorphic secondaries and tertiaries.

| $\theta$ | Surface | $R_x$ | $R_y$ | $K_x$ | $K_y$ |
|---|---|---|---|---|---|
| 7.50 | 2 | $-1.883\ 10^{+4}$ | $-3.717\ 10^{+4}$ | $-8.906$ | $-6.821$ |
| $\pm 0.1$ | 3 | $-1.782\ 10^{+3}$ | $-6.625\ 10^{+4}$ | $-54.872$ | $-0.811$ |
| 15.0 | 2 | $-1.007\ 10^{+4}$ | $-4.639\ 10^{+4}$ | $-30.822$ | $-9.234$ |
| $\pm 0.1$ | 3 | $-1.757\ 10^{+3}$ | $-3.446\ 10^{+4}$ | $-130.580$ | $-0.760$ |
| 22.5 | 2 | $-9.325\ 10^{+4}$ | $-2.210\ 10^{+3}$ | $-178.494$ | $-1.825$ |
| $\pm 0.1$ | 3 | $-4.78\ 10^{+4}$ | $-2.635\ 10^{+3}$ | $-265.011$ | $-0.557$ |

| $\theta$ | Surface | $A_r$ | $B_r$ | $C_r$ | $D_r$ |
|---|---|---|---|---|---|
| 7.50 | 2 | $-4.897\ 10^{-18}$ | $5.295\ 10^{-23}$ | $2.249\ 10^{-41}$ | $4.747\ 10^{-42}$ |
| $\pm 0.1$ | 3 | $3.057\ 10^{-15}$ | $1.587\ 10^{-22}$ | $2.775\ 10^{-33}$ | $5.382\ 10^{-35}$ |
| 15.0 | 2 | $-1.975\ 10^{-18}$ | $-1.738\ 10^{-22}$ | $1.224\ 10^{-40}$ | $1.323\ 10^{-38}$ |
| $\pm 0.1$ | 3 | $3.169\ 10^{-15}$ | $2.453\ 10^{-22}$ | $4.920\ 10^{-33}$ | $4.452\ 10^{-32}$ |
| 22.5 | 2 | $3.468\ 10^{-14}$ | $2.354\ 10^{-25}$ | $-7.862\ 10^{-31}$ | $5.793\ 10^{-45}$ |
| $\pm 0.1$ | 3 | $2.593\ 10^{-15}$ | $1.181\ 10^{-22}$ | $3.067\ 10^{-33}$ | $4.615\ 10^{-33}$ |

| $\theta$ | Surface | $A_p$ | $B_p$ | $C_p$ | $D_p$ |
|---|---|---|---|---|---|
| 7.50 | 2 | $-4.897\ 10^{-18}$ | $5.295\ 10^{-23}$ | $2.738\ 10^2$ | $-3.110$ |
| $\pm 0.1$ | 3 | $-7.374\ 10^1$ | $-2.044\ 10^1$ | $6.267\ 10^1$ | $5.574\ 10^{-1}$ |
| 15.0 | 2 | $-8.710\ 10^1$ | $-1.371$ | $-4.221\ 10^2$ | $-2.180$ |
| $\pm 0.1$ | 3 | $-8.188\ 10^1$ | $-2.093\ 10^1$ | $7.099\ 10^1$ | $8.589\ 10^{-1}$ |
| 22.5 | 2 | $3.487\ 10^{-1}$ | $-9.312$ | $2.962\ 10^{-1}$ | $2.229\ 10^{+2}$ |
| $\pm 0.1$ | 3 | $-6.383\ 10^1$ | $-2.103\ 10^1$ | $6.471\ 10^1$ | $-6.776\ 10^{-1}$ |

Examination of Figure (2) shows that the remaining aberrations are mostly high-order low-amplitude aberrations that could be corrected with adaptive optics. Adding adaptive optics to the system would improve its performance, extending its field of view and rendering it useful for high-resolution imagery. Given the high optical quality of liquid mirrors (Borra et al. 1993, Borra, Content & Girard 1993), the system could be adapted to diffraction-limited imagery with adaptive optics.

## 5. Conclusion

We have discussed a family of two-mirror correctors that can extend the field accessible to a fixed telescope such as a liquid mirror telescope. The performance of the corrector is remarkable since it yields excellent images in patches located within a field greater than 45 degrees. This performance is particularly interesting if one considers that optical telescopes are seldom used at zenith angles greater than 45 degrees beyond which the airmass increases rapidly, causing unacceptable degradation of performance from increased absorption and worsening seeing. This comment, based on a few hundred nights of observations on telescopes ranging from 5-m to 50 cm by one of us (EFB), is supported by a compilation (Benn & Martin 1987) of the statistics of the use of the Isaac Newton telescope at Las Palmas that finds that 94% of the observations were taken within a zenith distance of 50 degrees. In practice, a BMW corrector thus allows access to about half the sky accessible to a conventional telescope.

Once a sufficiently large accessible field is achieved, a fixed primary and movable correctors can actually yield a more efficient system than a classical tiltable telescope. This is because a classical telescope can only observe a field at a time, while a fixed primary with several correctors could access many widely separated fields simultaneously. This of course increases the complexity and cost of the telescope. The total number of simultaneously usable correctors will be limited by the requirement that their mechanical setups must not interfere. However, one can envision a corrector tracking a field North of the zenith at the same time as another one tracks in the South, while fixed correctors at the edge of the field carry out surveys in the driftscanning mode (Fig. (4)). This primary-sharing setup, allowing several research programs to be carried out simultaneously, is particularly attractive for very large telescopes, for which observing time is at a premium. We have designed several configurations where the secondary was not on the optical axis (as in Fig. (1)) so that it should be possible to avoid the mechanical incompatibility of multiple correctors that Figure (1) seems to indicate.

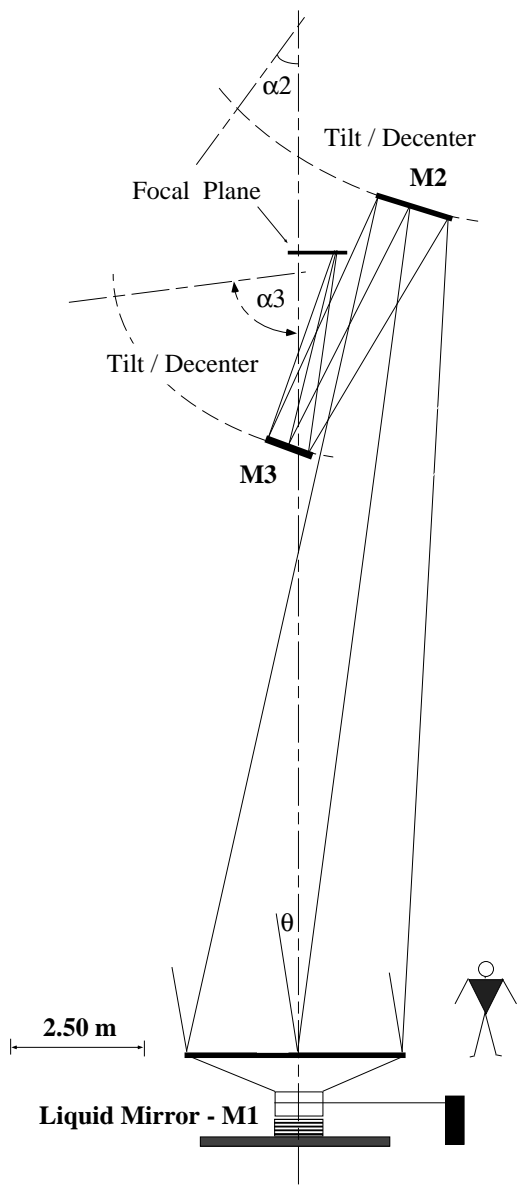

**Fig. 3.** It shows a schematic design of a 4-m diameter f/6 mirror and a BMW compact corrector observing at 15 degrees from the zenith.

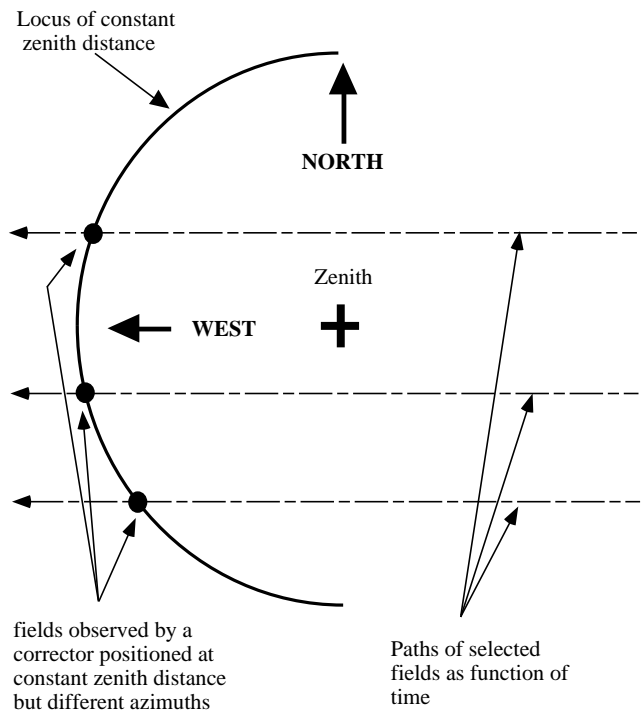

**Fig. 4.** It illustrates how a corrector set up for a given zenith distance $\theta$ (e.g. 22.5 degrees) can be used to observe objects passing within a field of view of $2\theta$ (e.g. 45 degrees) by moving the corrector at different azimuths but at a fixed zenith distance.

A BMW corrector is also attractive for a lunar-based telescope since weight is critical for such an instrument. We should expect that a fixed telescope equipped with a BMW corrector should be lighter than a conventional tiltable telescope, especially if the primary is a liquid mirror (Borra 1992).

The BMW corrector may not be the best practical corrector system. In particular, it may be possible to improve its performance with the addition of a third aspheric mirror.

We have confined our design to a 4-m mirror for discussion sake and, in the practical spirit of this article, to give an example for a telescope that could actually be presently built. However, this concept becomes particularly interesting for giant telescopes having diameters larger than 10-m. For those instruments, a fixed telescope may be the only practical design. Two telescopes, one built in Chile and the second in either Hawaii or the Mexican or United States desert, would access most of the sky.

*Acknowledgements.* This research was supported by grants from the Natural Sciences and Engineering Research Council of Canada and the Fonds FCAR of the province of Québec. G. Moretto was supported by a CAPES-Brazil postgraduate fellowship. We thank Mr. Jacques Racicot for estimating the cost of the silo. We thank Dr. H.R. Richardson for pointing out a mistaken assumption in a previous version of this article.